\documentstyle[12pt]{article}
\begin{document}
\newcommand{\newc}{\newcommand}
\newc{\ra}{\rightarrow} 
\newc{\lra}{\leftrightarrow} 
\newc{\beq}{\begin{equation}} 
\newc{\eeq}{\end{equation}} 
\newc{\barr}{\begin{eqnarray}} 
\newc{\err}{\end{eqnarray}}
\newc{\eps}{\epsilon}
\newcommand{\gl}{\lambda}
\newcommand{\PR}{{\it Phys. Rev. }}
\newcommand{\PRL}{{\it Phys. Rev. Lett.}}
\newcommand{\PL}{{\it Phys. Lett. }}
\newcommand{\NP}{{\it Nucl. Phys. }}
%%%%%%%%%%%%%%%%%%%%%%%%%%%%%%%%%%%%%%

$ $\hfill hep-th/9611035

~~\hfill IOA.06/96

~~\hfill NTUA 56-96\\
\vspace{1cm}

\begin{center}
{\bf
The $\mu$ -- term  in Effective Supergravity Theories.$^\dagger $
}%END \BF

\vspace*{.5cm}
{\bf 
G.K. Leontaris$^{(a)(1)}$ and N.D. Tracas$^{(b)(2)}$
} \\
\vspace*{.5cm}
{\it $^{(a)}$Theoretical Physics Division, Ioannina University}\\
{\it GR-451 10 Ioannina, Greece}\\
\vspace*{.3cm}
{\it $^{(b)}$Physics Department, National Technical University}\\
{\it GR-157 80 Zografou, Athens, Greece}\\
\end{center}

\begin{abstract}
The Higgs mixing term coefficient $\mu_{eff}$ is calculated in the scalar 
potential in supergravity theories with string origin, in a model independent
approach. A general low energy effective expression is derived, where
new contributions are included which depend on the modular weights
$q_{1,2}$ of the Higgs superfields, the moduli and derivative terms. We
find that in a class of models obtained in the case of compactifications 
of the heterotic superstring, the derivative terms are identically zero.
Further, the total $\mu_{eff}$-term vanishes identically if the sum of
the two modular weights $q_1+q_2$ is equal to two. Subleading 
$\mu$- corrections, in the presence of intermediate gauge symmetries
predicted in viable string scenarios, are also discussed.
\end{abstract}

\thispagestyle{empty}
\vfill
\noindent IOA 06/96\\
\noindent NTUA 56-96\\
\noindent October 1996

\vspace{.5cm}
\hrule
\vspace{.3cm}
{\small
\noindent
$^\dagger$Work partially supported by C.E.C Project CHRX-CT93-0319,
the TMR- program ERB FMR XCT 960090,  and
${\Pi}ENE\Delta$ 1170/95.\\
$^{(1)}$leonta@cc.uoi.gr\\
$^{(2)}$ntrac@central.ntua.gr

\newpage
%%%%%%%%%%%%%%%%%%%%%%%%%%

    In  the minimal supersymmetric standard model (MSSM), non-zero 
masses for the quarks and leptons require the existence of two higgs 
superfields $H_1,H_2$. In the effective superpotential of the model,
one of the higgs doublets couples to the up-type quarks,
while the second higgs provides with masses the down-type quarks 
and charged
leptons. If only these terms were present in the effective
superpotential, the latter is invariant under a Peccei-Quinn (PQ) 
symmetry
%%%%%%%%%
\cite{pq}
%%%%%%%%%
 which  finally  implies the existence of a higgs boson, the
 `electroweak axion', with zero bare mass
%%%%%%%%%%
\cite{ww}.
%%%%%%%%%%
A  way to avoid an unacceptably low mass for the axion
in the MSSM, is to introduce a  mixing term $\mu H_1H_2$
\cite{km,swdm}
where the mass parameter $\mu$ should be of the order of
the electroweak scale. The value of $\mu$ could be related
to the gravitino mass $m_{3/2}$ or arise from the vacuum
expectation value of a scalar component of a singlet field
$\phi H_1H_2\ra <\phi >H_1H_2$
%%%%%%%%%
\cite{km}.
%%%%%%%%
Nevertheless, the introduction of an explicit $\mu$-term in 
the theory generates a new hierarchy problem, since one
has to introduce a new scale in the theory, associated with
this mixing term. 

In the context of the $N=1$ effective supergravity theories,
which emerge as a low energy limit of a superstring theory,
it is possible to obtain an induced higgs mixing term\cite{agnt,kl}
due to the effects of a hidden sector. From the point of view
of string theories, the above features can be found in
models with a gauge group $G$ containing both an observable
and a hidden sector. In general, the observable part  has a
rank larger than that of the MSSM symmetry. Usually, $G$  breaks
{\footnote{In certain cases this breaking may occur radiatively\cite{lt0}} 
down to the Standard Model (SM) - gauge group at an intermediate
scale $M_X$, some two orders of magnitude below the string scale. 
A new mixing term for the Higgs fields responsible for the
intermediate symmetry breaking may also appear. In addition,
induced mixing terms from intermediate symmetry effects are
possible for the standard higgs doublets.

In this letter, we derive the $\mu$ - mixing terms in an effective 
supergravity theory with the generic stringy features described 
above. Taking into account these features and symmetries from
the string, one obtains an $N=1$ supergravity with the following
ingredients:
A real gauge invariant K\"ahler potential ${\cal K}$ which depends 
on the chiral superfields and moduli which are exact flat directions
of the scalar potential, and a superpotential ${\cal W}$ which is
a holomorphic function of the chiral superfields $Q_i$. 
The second derivatives of the K\"ahler potential determine the
kinetic terms of the various fields in the chiral supermultiplets,
while the Yukawa couplings appearing in the superpotential are
subject to string constraints.
The K\"ahler function ${\cal G}$ is defined
%%%%%%%%%%%%%  6
\cite{FKZ}
%%%%%%%%%%%%%%%%%%
\beq
{\cal G}(z,\bar z) = {\cal K}(z,\bar z) + \log{\mid {\cal W}(z)\mid}^2
\eeq
Denoting $z=(\Phi,Q)$,
where $\Phi$ stands for the dilaton field S and other moduli $T_i,U_i$
while Q stands for the chiral superfields, the superpotential ${\cal W}(z)$,
at the tree level, is given by
\beq
{\cal W}({\Phi},Q) = k_0 +
  \frac 1{2!} \mu_{ij}(\Phi)Q_iQ_j +
\frac 1{3!}\lambda_{ijk}({\Phi})Q_iQ_jQ_k +
 \cdots  \label{sp}
\eeq
where $k_0$ is in general a model dependent parameter\cite{fkpz}
which may have an explicit dependence on the moduli\cite{agnt}
 and $\{ \cdots \}$
stand for possible non - renormalizable contributions. Terms bilinear in the
fields
$Q_i$  refer in fact to a higgs  mixing term. At the perturbative level, due
to the analyticity of the superpotential and the presence of a
PQ - symmetry, the parameters $\mu$ and $\lambda$ do not depend
on the  dilaton field $S$.  Non-perturbative effects however,
may allow dilaton contributions to the superpotential of the form 
$\propto e^{-8\pi^2S}$, thus breaking the original PQ -
symmetry which allows only $S+\bar S$ dilaton combinations. 

In the following we will assume that the K\"ahler potential
${\cal K}$ can be expanded in inverse powers of the $(S+\bar S)$ 
fields with a tree level piece $ K_0 (T,\bar T)$  which takes the 
following general  form
%%%%%%%%  
\cite{agnt,kl}:
%%%%%%%%%%%%%%%%%%%%%
\begin{eqnarray}
{ K_0}(T,\bar T)= -\log(S+\bar S)
-\Sigma_nh_n\log(T_{n}+\bar T_{n}) + {\cal Z}_{i\bar j}Q_i{\bar Q}_{\bar j}
     +(\frac12{\cal M}_{ij}(T,\bar T)Q_iQ_j + c.c. )
\label{6}
\end{eqnarray}
where for simplicity $T$ represents all kinds of moduli except from
the dilaton field. 
Modular symmetries and K\"ahler transformations may be applied
to obtain the transformation  properties  of the tree level matrices
${\cal Z}$ and ${\cal M}$ as well as of the chiral fields and the 
superpotential\cite{kl,lt0,x}. 
We will soon see that one of the main sources of 
the induced $\mu$-term in the superpotential is the matrix ${\cal M}$
appearing in the K\"ahler function.

%%%%%%%%%%%%%%%%%%%%%%%%%%
In order to calculate the relevant 
contributions, we need the inverse 
K\"ahler metric $G_{I{\bar J}}^{-1}$, 
which in the basis chosen has 
the  block diagonal form
$ ({\cal K}_{S\bar{S}}^{-1},{\cal K}_{i\bar{j}}^{-1})$
where the subscripts denote differentiation with 
respect to the fields $z_I$
while ${\cal K}_{i\bar{j}}^{-1}$ is a 
$N+2\otimes N+2$ matrix with the indices
$i,j$ taking the values $0,1,...,N+1$
for the fields $T_{1,..N},H_1,H_2$
respectively. In particular, 
in the simplest case of the presence of only
one modulus $T$, ${\cal K}_{i\bar j}^{-1}$ is given by
\barr
{\cal K}_{i\bar j}^{-1}=\frac{1}{\Delta}
\left(\begin{array}{ccc}
 K_{1\bar 1}K_{2\bar 2}& -K_{0\bar 1}K_{2\bar 2}  &-K_{0\bar 2}K_{1\bar 1}\\
-K_{1\bar 0}K_{2\bar 2}& K_{0\bar 0}K_{2\bar 2}-K_{0\bar 2}K_{2\bar 0}
&K_{1\bar 0}K_{0\bar 2}\\
-K_{2\bar 0}K_{1\bar 1}&K_{0\bar 1}K_{2\bar 0}&
  K_{0\bar 0}K_{1\bar 1}-K_{0\bar 1}K_{1\bar 0}\\
\end{array}\right)\label{matrixK}
\err
where $\Delta$ denotes the determinant of the ${\cal K}_{i\bar{j}}$ matrix.
The extension to $N$-moduli, it straightforward.
To proceed further, we find it convenient to define the 
following covariant $\mu$-derivatives,
\barr
{\cal D}_T{\tilde\mu}_{ij}&=&\partial_T\mu_{ij} +
{\cal W}\partial_T{\cal M}_{ij}\\
{\cal D}_T\bar{\tilde\mu}_{\bar i\bar j}&=&
\partial_T\bar\mu_{\bar i\bar j} +
\bar{\cal W}\partial_T{\bar{\cal M}}_{\bar i\bar j}
\err
and the combination
\beq
{\tilde\mu}_{ij} = \mu_{ij} +{\cal W}{\cal M}_{ij}
\eeq
The part of the effective scalar potential related to the
supersymmetry breaking effects is given by 
\beq
V_F = e^{\cal G}\left(
{\cal G}_{\bar I}{\cal G}_{I\bar J}^{-1}{\cal G}_J - 3\right) 
+\cdots
\label{vf}
\eeq  
where $\{\cdots\}$ represent $D$-term contributions.
Assuming a form of the K\"ahler potential and the superpotential
dictated from modular symmetries,
we can now define through (\ref{vf}) the boundary conditions for soft
mass terms as well as the induced higgs mixing. As stressed in the
introduction, 
any low energy effective supersymmetric field theory contains in
its massless spectrum at least two higgs fields associated with
the standard two doublets of the MSSM and the existence of
 a higgs mixing term, --  $\mu$-term --
in theories of two higgs doublets is  necessary.
In effective quantum field  theories arising from 
the heterotic string, the form of the 
K\"ahler potential may provide such terms in the effective
superpotential. 

Thus, in the K\"ahler function  we will assume the 
existence of a higgs mixing term of the form ${\cal M}_{ij}H_iH_j$
where  ${\cal M}$ depends on the moduli $(T_n,\bar T_n)$.
An explicit higgs mixing term (the $\mu$-term) may also exist in the 
original superpotential of the model. 
  The most general form of the tree level superpotential 
arising in the theories under consideration, has been written
in Eq.(\ref{sp}). As explained above, we will restrict our analysis
in cases where the tree level Yukawa couplings of the superpotential
${\cal W}$  are functions of the moduli $T_n$, i.e., 
$\mu_{ij}(T),\lambda_{ijk}(T)$ but at the tree level,  
they do not depend on the dilaton $S$. For a more
involved situation however, in a final example
we will allow the possibility of the existence
of an `unobservable' phase $\varphi (T,\bar T)$
for the case  of the $\mu$ - tree level term, which could 
in principle depend on  $T$ and $\bar T$  moduli. Such a phase can
be justified from the transformation properties under 
modular invariance  of
the physical mixing mass in certain compactifications
of the heterotic string theory\cite{agnt}.  Then,
we will soon see that due to the presence of induced $\mu$ 
contributions involving the derivatives of higgs mixing 
mass terms, such a phase will manifest itself in the 
effective $\mu$-term. Finally, due to the possible existence
of the intermediate symmetry breaking, new threshold effects 
can arise and in principle should not be ignored. 

Under the above assumptions, we calculate the quantities
involved in the effective potential including also terms
proportional to the vacuum expectation values (vevs) of the higgs.       
The various kinds of derivatives which can arise are the following
\barr
G_0&=&-\frac{h}{\tau}-\sum_i\frac{q_i}{\tau^{1+q_i}}H_i{\bar H}_{\bar i} +
     \frac{1}{2}{\cal W}^{-1}{\cal D}_T{\tilde\mu}_{ij}H_iH_j+
     \frac{1}{2}{\bar{\cal W}}^{-1}{\cal D}_T\bar{\tilde\mu}_{\bar i\bar j}
                                  {\bar H}_{\bar i} {\bar H}_{\bar j} 
\label{G0} \\
G_{\bar 0}&=&-\frac{h}{\tau}-
\sum_i\frac{q_i}{\tau^{1+q_i}}H_i{\bar H}_{\bar i} +
     \frac{1}{2}{\cal W}^{-1}{\cal D}_{\bar T}{\tilde\mu}_{ij}H_iH_j+
\frac{1}{2}{\bar{\cal W}}^{-1}{\cal D}_{\bar T}\bar{\tilde\mu}_{\bar i\bar j}
                                  {\bar H}_{\bar i}{\bar H}_{\bar j} \\
G_i&=&\tau^{-q_i}{\bar H}_{\bar i} +{\cal W}^{-1}{\tilde\mu}_ {ij}H_j\\
G_{\bar j}&=&\tau^{-q_j}{H}_j+
    {\bar{\cal W}}^{-1}\bar{\tilde\mu}_{\bar i\bar j} {\bar H}_{\bar i }
\err
where $\tau=T+\bar T$ and $q_i$ are the modular weights of the corresponding
higgs field.
To obtain the inverse metric we also need the elements of $K_{I\bar J}$ matrix 
which are given by
\barr
K_{0\bar 0}=\frac h{\tau^2}&+&
            \sum_i\frac{q_i(q_i+1)}{\tau^{2+q_i}}H_i{\bar H}_{\bar i}
\nonumber\\
&+&\frac 12\partial_{\bar T}\left[{\cal W}^{-1}{\cal D}_T\mu_{ij}H_iH_j
+\bar{{\cal W}}^{-1}{\cal D}_T\bar{\mu}_{ij}{\bar H}_{\bar i}{\bar
H}_{\bar j}\right]
        \err
 \barr
 K_{0\bar i}=&-&\frac{q_i}{\tau^{1+q_i}}H_i\nonumber\\
 &+&{\bar{\cal W}}^{-1}
\left(\partial_T\bar{\tilde\mu}_{\bar i\bar j}-
 \frac 12\bar{{\cal W}}^{-1}\bar{\tilde\mu}_{\bar i\bar j}
 \partial_T\bar{\mu}_{\bar k\bar\ell}\bar{H}_{\bar k}
 \bar{H}_{\bar\ell}\right)\bar{H}_{\bar j}\err
 
 \barr
 K_{i\bar j}=\frac{1}{\tau^{q_i}}\delta_{i\bar j}
 \err
 while the determinant is given by
 \barr
 \Delta&=&\frac{1}{\tau^{2+q_1+q_2}}
           \left[h+\sum_iq_i(q_i+1)\tau^{-q_i}H_i\bar H_{\bar i}
\nonumber\right.\\
       &+&\frac 12\tau^2
       \partial_{\bar T}\left(W^{-1}{\cal D}_T\mu_{ij}H_iH_j+
 %\left.
 \bar{W}^{-1}{\cal D}_T\bar\mu_{\bar i\bar j}
                                   \bar H_{\bar i}\bar H_{\bar j}
 \right)
  \label{deter}
  \\
 &+&\tau^{q_2}
 \left(-q_2\tau^{-q_2}H_2+
 \tau\partial_T
 \left({\bar{\cal W}}^{-1}\bar{\tilde\mu}_{\bar i\bar 2}{\bar H}_{\bar i}
 \right)\right)
 \left(-q_2\tau^{-q_2}{\bar H}_{\bar 2}+
 \tau\partial_{\bar T}
 \left({\cal W}^{-1}\tilde\mu_{2j} H_j
 \right)
 \right)
 \nonumber \\
 &+&\tau^{q_1}
 \left.
 \left(-q_1\tau^{-q_1}H_1+
 \tau\partial_T
 \left({\bar{\cal W}}^{-1}\bar{\tilde\mu}_{\bar i\bar 1}{\bar H}_{\bar i}
 \right)\right)
 \left(-q_1\tau^{-q_1}{\bar H}_{\bar 1}+
 \tau\partial_{\bar T}
 \left({\cal W}^{-1}\tilde\mu_{1j} H_j
 \right)
 \right)
 \right]\nonumber
 \err
In the presence of higgs fields with vevs not very
far from the unification point, there are in principle, 
numerous  mixing terms arising from all combinations
of light and heavy higgs fields through 
 the quantity  $G_{\bar J}G_{J\bar I}^{-1}G_I $.
However, in the following we will assume that 
the intermediate gauge symmetry breaks down to
the standard gauge group at a scale at least one or two
orders of magnitude bellow $M_{string}$. The 
possible vev - dependent $\mu$-contributions depend
quadratically on these vevs, thus they are rather suppressed.
On the contrary, there exist vev - independent contributions
which are of the order of the gravitino mass $m_{3/2}$, and/or 
the possibly existing explicit $\mu$ - mass term of the 
tree level superpotential.  Obviously, since  these  terms 
are independent of the large higgs vev's, it turns out that
they are present even in the absence of any intermediate symmetry.

In the following we present first the vev-independent contributions
and  show their origin. It is enough for the moment to concentrate
on the SM-higgs doublets.  First we approximate ${\Delta}\sim
\frac{h}{\tau^{2+q_1+q_2}}$, while we assume a
 single pair of higgs fields. In this case we will simplify
our notation by the replacement $\mu_{ij}\ra \mu_{12}$
or even simply $\mu$.
Starting from the diagonal terms  $G_{\bar I}G_{I\bar I}^{-1}G_I $,
where $I=T,H_1,H_2$,  we obtain for $I=T$,
\beq
G_{\bar T}G_{T\bar T}^{-1}G_T \ra
- \frac 1{{\cal W}}(T+\bar{T})\left\{ D_{ T}
+ D_{\bar T}\right\} 
\tilde{\mu}_{12}\label{c1}
\eeq
while two more contributions result from the diagonal terms with respect 
to the derivatives of the two higgs fields $H_{i=1,2}$, namely 
\beq
\sum_i G_{\bar i}G_{i\bar i}^{-1}G_i {}\ra
     2\frac{\tilde{\mu}_{12}}{{\cal W}}
\label{c2}
\eeq
The  terms in Eqs(\ref{c1},\ref{c2}) are the same with those obtained 
in previous works\cite{agnt,kl} and constitute the Yukawa coupling of
the corresponding fermion  mass.
Now, in the scalar potential the corresponding soft parameter receives
additional contributions from off-diagonal terms $G_{\bar I}G_{I\bar J}^{-1}G_J $
where $I\not= J$. In particular, it can be easily seen  
that these contributions are obtained from the two terms
$G_{\bar T}G_{T\bar i}^{-1}G_i $, $i=1,2$. Two types
of terms may arise here. The first one is directly
proportional to the combination of $\tilde\mu$ multiplied
by the modular weight $q_i$ of the corresponding higgs field
$H_i$. There is a second term common in both ($i=1,2)$ terms 
which depends on the properties of the quantity  $\tilde\mu$
under differentiation with respect to the moduli. More 
explicitly for the first  of the two contributions we have
%  {\it I}) 
 \barr
\sum_i G_{\bar T}G_{T\bar i}^{-1}G_i\ra  -(q_1+q_2)
\frac{{\tilde\mu_{12}}}{{\cal W}}
\label{new1}
 \err
where the
normalization of the fields has been taken into account. The second 
type is proportional to the covariant $\mu$-derivative, i.e.,\\
% {\it II}) 
\beq
 \frac 2{{\cal W}}\left[(T+\bar T)D_{\bar T}\right]
{\tilde\mu_{12}}
\label{new2}
\eeq 
A third contribution similar to the second
is also possible, however this is proportional to the  
$\mu_{12}\upsilon_1\upsilon_2\partial_T{\cal M}_{12}$,
where $\upsilon_i$ is the vev of the corresponding higgs,
and is assumed to be small.
The remarkable fact however, is that even if the higgs vevs
are sent to zero and no intermediate scale exists, these 
new mixing terms from off-diagonal $K_{I\bar J}^{-1}$-elements
contribute substantially to the  $\mu$-term. In particular in the
large  radius limit, i.e., for
large values of the moduli $T > 1$, the contribution
(\ref{new2}) might be  significant as it is
proportional to $T+\bar T = 2Re{T}$ and should not be ignored.
We may conclude that, although the analysis above is done for rather
general effective supergravity models, the parameters entering
the $\mu$-formula are rather constrained.
Indeed, starting from the second term in (\ref{new1}), it is a remarkable 
fact that only the sum of the two Higgs modular  
weights $q=q_1+q_2$ enters in the $\mu$ expression. Although the
$q_i$  themselves  are model dependent, the value of $q$ however, 
could be constrained from general requirements. For example, certain
constraints can be put on $q$\cite{lt0,ekn} from the
transformation properties of the superpotential terms.

Let us now collect the above contributions into an effective
higgs mixing mass term. For practical purposes it is useful
to simplify the above formulae and keep the leading terms. 
With the definition,
\begin{eqnarray}
\mu_{sim}(T,\bar{T})\equiv \frac{\mu}{c} + {\cal M}
\end{eqnarray}
with $c$ being a numerical value associated with the
vacuum expectation value of the superpotential,
\begin{eqnarray}
c=<{\cal \mid W\mid}>=e^{-<{\cal K}>/2} m_{3/2} 
\end{eqnarray}
we summarize our results in the following simple formulae
\begin{eqnarray}
G_{\bar 0}K^{-1}_{0\bar 0}G_0&\rightarrow&- h \tau^{-(1+q)}
                           (\partial_T+\partial_{\bar T})\mu_{sim}
\nonumber\\
G_{\bar 1}K^{-1}_{1\bar 1}G_1 + 
G_{\bar 2}K^{-1}_{2\bar 2}G_2&\rightarrow& 2 h \tau^{-(2+q)}
                                                  \mu_{sim} 
\nonumber\\
 G_{\bar 0}K^{-1}_{0\bar 1}G_1 
+G_{\bar 0}K^{-1}_{0\bar 2}G_2&\rightarrow&- q h \tau^{-(2+q)}
                                                 \mu_{sim}
\label{simplemus}\\
 G_{\bar 1}K^{-1}_{1\bar 0}G_0
+G_{\bar 2}K^{-1}_{2\bar 0}G_0&\rightarrow&2 h \tau^{-(1+q)}
                                         \partial_{\bar T}\mu_{sim}
\nonumber\\
G_{\bar 1}K^{-1}_{1\bar 2}G_2,\quad 
G_{\bar 2}K^{-1}_{2\bar 1}G_1&\rightarrow&0
\nonumber
\end{eqnarray}
Adding all the above terms and dividing by the determinant
 $ h \tau^{-(2+q)}$ we arrive at our final result for the
leading part of the low energy coefficient of
the effective $\mu$-term,
\beq
\frac{\mu_{eff}}{m_{3/2}}  =  \left\{1-\tilde{q}-
 {\rm Re}T
 ({\partial_T-\partial_{\bar T}})\right\}\mu_{sim}(T,\bar T)
\label{mueff}
\eeq
with $\tilde{q}=\frac{q_1+q_2}{2}$.
This formula can be further simplified in models where $\mu$ and ${\cal M}$
parameters are having simple and well defined properties under the
modular transformations. Consider in particular the case where $\mu$
is a constant whilst ${\cal M}(T,\bar{T})$ has a scaling property
under the $T$ and $\bar{T}$ derivatives\cite{agnt}, i.e.,
$$(T+\bar{T})\partial_{T/\bar{T}}{\cal M}(T,\bar{T})
={\cal M}(T,\bar{T})$$ 
In this case, the derivative term in (\ref{mueff}) vanishes and
the formula takes the simple form
\beq
\mu_{eff}' = (1-\tilde q)(e^{<{\cal K}>/2}\mu + m_{3/2}{\cal M})
\label{mueff'}
\eeq
We should point out here, that under the above assumptions 
we can see from (\ref{mueff'}) that there exists a possibility where
the presence of the higgs mixing term ${\cal M}$ 
in the K\"ahler function does not
imply an effective $\mu$-term in the low energy potential, namely
when $q_1+q_2 = 2$. In fact, as we will see in our example, this is
the case of a class of  string models obtained in the (2,2) 
compactifications of the heterotic superstring provided that
the explicit $\mu$-term in the superpotential 
is either constant or absent. 

As we have explained above, in the case of the intermediate symmetry
additional terms can play a role in the mixing of
the higgs fields involved in the symmetry breaking. 
The sub-leading $\mu$-contributions  are proportional  to
$\upsilon_i\upsilon_j$ and have the 
highest negative power of $\tau$. We note
that such terms can come also from the expansion of $\Delta$.
Indeed, from the sub-leading terms of the matrix (\ref{matrixK}) and
the leading, vev-independent, term of $\Delta$, we get the following
contribution to the the $\mu$-term 
(for simplicity we assume $\upsilon_i=\upsilon_j=\upsilon$)
\[ \tau^{-\tilde{q}}h^{-1} 2(q_1q_2+\tilde{q})\upsilon^2   \]
while expanding the $\Delta$ and taking the sub-leading terms we get
\[ \tau^{-\tilde{q}}h^{-1} 2(-\tilde{q})\upsilon^2   \]
Adding up we get the total sub-leading contribution
\beq
 \tau^{-\tilde{q}}h^{-1} 2q_1q_2\upsilon^2  \label{submu}
\eeq 
For example if $\tilde{q}\equiv (q_1+q_2)/2$ is around 3, 
then this contribution is a 10\% correction
to the leading term, assuming $\tau\sim 0.1$ and $\upsilon\sim 0.01$,
while they are suppressed in the large radius limit.

%%%%%%%%%%%%%%%%%%  Kahler

As an application of the above procedure, 
we consider the general form of the K\"ahler function
\beq
K(T,\bar T,Q_i,\bar Q_i) = 
- \sum_i h_i\log\left[\prod_n(T_n+\bar T_n)^{q_i^n}-
\hat Q_i\hat{\bar Q}_i\right] 
\eeq
where the $\hat{Q}_i$ denote fields which, in general,
correspond to linear combinations of the eigenstates.
Expanding the logarithm in terms of the eigenstates, one gets
\beq
K(T,\bar T,Q_i,\bar Q_i) = -\sum_{n,i}
 h_iq_i^n \log\left(T_n+\bar{T}_n\right)
    + {\cal Z}_{i\bar j}Q_i\bar{Q}_i+
\frac12{\cal M}_{ij}Q_iQ_j+ \cdots
\eeq
where the matrices ${\cal Z}\, ,{\cal M}$ are proportional to 
$\prod_n(T_n+\bar T_n)^{-q_i^n}$.

Our example is a generalization of the K\"ahler forms obtained
\cite{agnt} in (2,2) compactifications of the heterotic 
superstring. Nevertheless, it can be easily seen that as far as 
we work at the tree level approximation, the 
approximated  K\"ahler  potential has definite properties
under the group of modular transforms\cite{lt0} and 
it is the same in both cases. We will present here a specific example in
order to see how a matrix ${\cal M}_{ij}$ may arise.
Consider for example the case of two moduli $T,U$ and
the fields $Q=A+\bar B$, $\bar{Q}=\bar{A}+B$ of 
ref.\cite{agnt}, where $A,B$ belong to $27$ and $\bar{27}$
of $E_6$. The K\"ahler function reads
\beq
K = -\log\left\{(T+\bar T)(U+\bar U)-
         (A+\bar B)(\bar A + B)\right\}
\eeq
where $A,B$ are identified with the higgs fields. Expanding in 
terms of the latter, one gets at first order the higgs mixing
 term ${\cal M}$  
\beq
{\cal M} = \frac{1}{(U+\bar U)(T+\bar T)}
\label{Khmt}
\eeq
which has the same form as in the general case above.
Let us now return to our $\mu$-formulae. Due to the properties of
the K\"ahler function and assuming $\mu$ constant, we conclude
that  $\mu_{eff}$ is given by (\ref{mueff'}). Moreoever, in
a class of  models compactified on an orbifold
the untwisted $A,B$ fields associated with the higgses, transform
as modular forms of weight 1, thus the sum $q_1+q_2$ is equal to two
and the total $\mu_{eff}$ vanishes identically. Thus, 
even if the K\"ahler potential contains a higgs mixing term
of the form (\ref{Khmt}), due to intriguing cancellations, it is
not possible to generate an effective $\mu$-term in the scalar 
potential within this class of 
orbifold string constructions unless an explicit - moduli
dependent - $\mu$  term
is present in the superpotential. This is the case of the 
particular model discussed in\cite{agnt}.

Furthermore, consider a more general case where 
the $\mu$ parameter of the superpotential 
depends on a phase factor of the form 
\beq
\mu (T,\bar T) = \mu_0(T)
\left(\frac{\imath c T +d}{-\imath c\bar{T}+d}\right)^{1/2}
\eeq
while assume a scaling property for  $\mu_0(T)$,
 i.e., $(T+\bar T)\partial_T\mu_0(T)=\mu_0(T)$. 
The interesting point to note here
is  that  this phase  will have an observable
effect through the derivative part in the $\mu_{eff}$ 
formula (\ref{mueff}).  In fact this term will give a
contribution 
\beq
(T+\bar T) (\partial_T-\partial_{\bar T})\mu{(T,\bar T)}
= \left\{1+ 2\left(\frac{Re T}{\mid\imath T+ d/c\mid}\right)^2
\right\}\mu{(T,\bar T)}
\eeq
which is proportional to $\mu$, up to a factor whose existence
is due to the $\mu$-phase.

In conclusion, in the context of effective supergravities 
characterised by properties of compactified heterotic string 
theories, we have derived a  general form of the effective
higgs mixing term $\mu_{eff}$ of the low energy effective
scalar potential. Using a  gauge invariant
form of the K\"ahler function, constrained by modular symmetries,
 we find additional contributions to $\mu_{eff}$. To leading 
order, these are found to depend on a specific combination 
$\mu_{sim}$ of the higgs mixing -- moduli-dependent -- matrix
 ${\cal M}$ of the K\"ahler potential and the possible 
 $\mu$-term coefficient of the superpotential as shown in 
formula (\ref{mueff}). Thus, all possible sources 
can be classified in the following two categories:
{\it i}) a term directly proportional to this
combination  with a proportionality factor $1-{\tilde{q}}$
where ${\tilde{q}}$ is half the sum of the modular weights
of the two higgs fields breaking the symmetry, and
{\it ii}) a derivative term on $\mu_{sim}$ with respect to 
the moduli $T,\bar{T}$. 

We discussed models with properties dictated by modular symmetries
where some of these contributions vanish.  
We further examined cases corresponding to models with
intermediate symmetry breaking scales which are not
far from the string unification point. There, in addition
to the above contributions there are vev-dependent terms
which could be important in specific regions of the
${\tilde{q}},T$, vev - parameters.

\vspace*{1cm}

We would like to thank C. Kounnas for helpful discussions.
N.D.T. wishes to thank the Theory Division of CERN for kind
hospitality. G.K.L. would like to thank J.D. Vergados for
a stimulating discussion and I. Antoniadis for useful
comments.

\end{document}